# Fe-resonant valence band photoemission and oxygen NEXAFS study on $La_{1-x}Sr_xFe_{0.75}Ni_{0.25}O_{3-\delta}$

*Selma Erat[1,2,a], Hiroki Wadati[3], Funda Aksoy[4,5], Zhi Liu[5],*

*Thomas Graule[1,6], Ludwig J. Gauckler[2], Artur Braun[1,b]*

[1]Laboratory for High Performance Ceramics
Empa. Swiss Federal Laboratories for Materials Science & Technology
CH - 8600 Dübendorf, Switzerland

[2]Department for Materials, Nonmetallic Inorganic Materials,
ETH Zürich, Swiss Federal Institute of Technology
CH - 8037 Zürich, Switzerland

[3]Department of Physics and Astronomy and AMPEL
University of British Columbia,
Vancouver, British Columbia V6T 1Z4, Canada

[4]Department of Physics, Cukurova University
TR - 01330 Adana, Turkey

[5]Advanced Light Source
Ernest Orlando Lawrence Berkeley National Laboratory,
Berkeley, California 94720, USA

[6]Technische Universität Bergakademie Freiberg
D-09596 Freiberg, Germany

[a] Corresponding authors: selma.erat@empa.ch , artur.braun@alumni.ethz.ch,
Phone: +41 44 823 4850, Fax: +41 44 823 4150.

ABSTRACT

Iron resonant valance band photoemission spectra of Sr substituted $LaFe_{0.75}Ni_{0.25}O_{3-\delta}$ have been recorded across the Fe 2p – 3d absorption threshold to obtain Fe specific spectral information on the 3d projected partial density of states. Comparison with $La_{1-x}Sr_xFeO_3$ resonant VB PES literature data suggests that substitution of Fe by Ni forms electron holes which are mainly O 2p character. Substitution of La by Sr increases the hole concentration to an extent that the $e_g$ structure vanishes. The variation of the $e_g$ and $t_{2g}$ structures is paralleled by the changes in the electrical conductivity.

**Introduction**

The electronic structure of defect regulated materials is very important for understanding their functional relationship such as transport properties. The electrical conductivity of oxides with perovskite structure depends on their valence band characteristic [1]. $La_{1-x}Sr_xFeO_{3-\delta}$ is a parent compound for solid oxide fuel cell cathodes [2], and the conductivity particularly at high temperature is an important quantity for its functionality as a cathode. Sr substitution of $LaFeO_3$, which is an insulator, forms electron holes which are associated with nominally tetravalent Fe having the electronic structure of $3d^5\underline{L}$, $\underline{L}$ denoting the electron hole from the oxygen ligand. Substitution by 50% ($Fe^{3+}/Fe^{4+}$ = 1/1) warrants that the electron hopping across the $Fe^{3+}$–O–$Fe^{4+}$ superexchange unit occurs with maximum probability. It has been shown that Fe substitution of $La_{1-x}Sr_xFeO_{3-\delta}$ on the B-site by Ni [3] or Co causes an additional significant improves the conductivity, with a conductivity maximum at temperatures around 650 K to 750 K, rendering them attractive for intermediate temperature proton conducting ceramic fuel cells [4]. The structural and transport properties of $La_{1-x}Sr_xFe_{1-y}Ni_yO_{3-\delta}$ has been studied for different stoichiometry (x,y) [5-7]. It has also been explained how the electronic structure evolves from charge transfer-type insulator $LaFeO_3$ (2 eV band gap) to the "oxygen-hole metal" $SrFeO_3$ [8] using *in situ* photoemission spectroscopy (PES) on thin films. Out of the rare-earth nickel oxides, $LaNiO_3$ is the only metallic member [9]. Chainani et al. [10] showed a systematic X-ray photoemission study on bulk LSFO. However, the understanding of Ni and Sr substitution effects on electronic structure especially for occupied states of Fe is still lacking.

We employ Fe resonant VB PES in order to reveal the influence of the Ni-substitution and Sr-substitution with respect to the Fe, which is the main player for the electrical conductivity via the $Fe^{3+}$–O–$Fe^{4+}$ superexchange unit.

The A-site substitution of $LaFeO_3$ by Sr ($La_{1-x}Sr_xFeO_3$) and its impact on the electronic conductivity is well understood. In the oxygen near edge x-ray absorption fine structure (NEXAFS) spectra this substitution manifests in a so called hole doping state with $e_g\uparrow$ symmetry as the leading peak in the O 1s pre-edge, preceding the two $t_{2g}\downarrow$ and $e_g\downarrow$ orbital symmetry peaks from the hybridized O2p-Fe3d states. Interestingly, B-site substitution with Ni, too, shows such extra peak. The question naturally arises as to which role Fe and the Ni play for the corresponding spectra.

**Experimental Section**

High purity precursors ($La_2O_3$ >99.99 %, $SrCO_3$ 99.9 %, $Fe_2O_3$ >99.0 %, and NiO 99.8 %, purities given in %) were mixed in stoichiometric proportions, ball milled for 24 hours, calcined for 4 h at 1473 K and for 12 h at 1673 K so as to obtain $La_{1-x}Sr_xFe_{0.75}Ni_{0.25}O_{3-\delta}$ (LSFN) with x=0, 0.50, 0.75 [9]. X-ray powder diffractograms confirmed that the samples had rhombohedral symmetry (space group: R-3c, space group no: 167), with a barely noticeable contamination of a tetragonal phase [10]. The 4-point DC conductivity shows the same trend like in ref. [9]: the conductivity increases with Sr doping and at x=0.50 shows a maxima and then starts to decrease. For valence band photoemission measurements the fine powders were uniaxially pressed into discs (~Ø 13 mm) at 40 KN and then sintered at 1673 K for 12 h. Photoemission and x-ray absorption spectra were recorded at end station [11] at beamline 9.3.2 of the Advanced Light Source, Lawrence Berkeley National Laboratory. The energy resolution of the beamline is $E/\Delta E = 3000$ [12]. Spectra were recorded in the maximum transmission detection mode. During measurements the base pressure of the main chamber was maintained at $10^{-9}$ Torr. The resonant valence band photoemission spectra for Fe metal were measured using photon energies between 704 eV and 716 eV, leading up to the $2p_{3/2}$ resonance energies. Near edge x-ray absorption fine structure (NEXAFS) spectra were measured at beamline 8.0.1 in total electron yield (TEY) mode at 300 K.
I did never grind the precursor chemicals.

**Results and Discussion**

Figure 1(a), (b), and (c) show the valence band spectra of $La_{1-x}Sr_xFe_{0.75}Ni_{0.25}O_{3-\delta}$ for x = 0.0, 0.50, and 0.75 obtained by tuning the excitation photon energy across the Fe 2p → 3d absorption threshold. The expanded valence-band spectra at Fe 2p-3d on-resonance (710 eV) are shown in Figure

1(d). Figure 2 shows the normalized Fe $L_3$ ($2p_{3/2}$) and $L_2$ ($2p_{1/2}$) x-ray absorption spectra for x=0 and x=0.50, marked with the energy positions, at which the resonant PES data were recorded. The Fe L edge is split into an $L_3$ and $L_2$ multiplet due to the core hole spin orbit coupling. The $L_3$ and $L_2$ edges are additionally split by the crystal field into peaks from transitions that arise from states with $t_{2g}$ and $e_g$ orbital symmetry. Visual inspection of the two spectra suggests that the sharp multiplet for x = 0 corresponds to Fe with predominant $Fe^{3+}$ valence, whereas the somewhat more diffuse multiplet for x = 0.5 corresponds to Fe with a mixed valence between $Fe^{3+}$ and $Fe^{4+}$, particularly because of the relatively less intense $t_{2g}$ structure in the $L_3$ edge. Comparison of simulated spectra based on atomic multiplet theory and measured spectra indeed shows that a very good match is obtained when for x = 0 Fe is entirely $Fe^{3+}$, and for x = 0.5 Fe is to equal amounts in $Fe^{3+}$ and in $Fe^{4+}$, all of which in the high spin state [9].

Because the spectra are normalized to the photon flux, the resonant enhancement of the spectral intensity at 710 eV (on-resonance), which reflects the 3d partial density of states of $La_{1-x}Sr_xFe_{0.75}Ni_{0.25}O_{3-\delta}$, becomes immediately obvious. It is instructive to enter the discussion of the spectral features of $La_{1-x}Sr_xFe_{0.75}Ni_{0.25}O_{3-\delta}$ by referring to the previously discussed VB PES spectra of $La_{1-x}Sr_xFeO_{3-\delta}$ [13]. The VB spectra are known to consist of four principal structures, i.e. an $e_g$ state (A) near the Fermi energy, a state with $t_{2g}$ symmetry (B), the Fe 3d – O 2p bonding state peak (C), and a charge transfer satellite (D) [13]. The $La_{1-x}Sr_xFe_{0.75}Ni_{0.25}O_{3-\delta}$ samples studied here are polycrystalline sintered bodies with a finite porosity. Such samples have a large surface-to-volume ratio than well finished single crystals. It is known that $La_{1-x}Sr_xFeO_{3-\delta}$ surfaces are slightly understoichiometric and oxygen deficient, thus providing the iron atoms near the surface a less than sixfold octahedral coordination. This may be a reason why the VB structures near $E_F$, particularly the $t_{2g}$ and $e_g$ structures are less well developed in our sintered polycrystalline samples than in single crystals.

The general comparison of the VB PES spectra for $La_{1-x}Sr_xFeO_{3-\delta}$ and $La_{1-x}Sr_xFe_{0.75}Ni_{0.25}O_{3-\delta}$ shows that the $e_g$ peak (A) is virtually absent in $La_{1-x}Sr_xFe_{0.75}Ni_{0.25}O_{3-\delta}$, whereas the $t_{2g}$ peak (B) is noticeable at around 3.5 eV and the O 2p bonding peak well pronounced at around 7.5 eV, as well as the satellite at around 11.5 eV. Closer inspection of the on-resonance with hν = 710 eV in Figure 1(d), which displays the features magnified in the energy range from -1.0 eV to 4.0 eV, shows that the intensity of the spectrum for x=0 is enhanced for 0.20 eV < E < 1.70 eV in comparison to the spectra for x = 0 and x = 0.75, and thus identify the structure at around 0.9 eV as from the $e_g$ symmetry derived transition corresponding to the one known for $LaFeO_3$ [13]. Upon substitution of Sr by 50% and by 75%, this $e_g$ (A) structure is not noticeable anymore. The next observation is that the energy position of the $t_{2g}$ (B) structure seems to move towards higher binding energy, i.e. from 3.5 eV for x = 0 to just below 4.0 eV for x = 0.5, and also for x = 0.75. This is in contrast to observations made on $La_{1-x}Sr_xFeO_{3-\delta}$. It appears also that the relative intensity of the $t_{2g}$ peak decreases when x is increasing. Thus, the general picture is that spectral intensity moves way from the Fermi energy $E_F$ with increasing substitution parameter x, as is particularly evidenced for the peaks due to transitions from states with the $e_g$ and $t_{2g}$ orbital symmetry. For $La_{1-x}Sr_xFeO_{3-\delta}$, the intensity decrease of the $e_g$ (A) structure upon Sr substitution was rationalized by transfer of doped electron holes towards the Fe $e_g$ band [7].

Since states with $e_g$ orbital symmetry correspond to Fe in octahedral coordination with the oxygen ions, the diminishing of the $e_g$ state can thus be interpreted as a deviation from octahedral coordination. We therefore conclude that substitution of Fe by Ni, too, leads to formation of electron holes which move to the Fe $e_g$ bands, and therefore Ni substitution has the similar effect on the depletion of the $e_g$ structure like Sr-substitution. In addition, as evidenced in our spectra, the established trend that substitution of La by Sr increases the doped holes concentration in $La_{1-x}Sr_xFeO_{3-\delta}$, is also recovered in our $La_{1-x}Sr_xFe_{0.75}Ni_{0.25}O_{3-\delta}$ samples: The hole states formed upon Sr substitution in $La_{1-x}Sr_xFeO_{3-\delta}$ and LSFN deplete the $e_g$ states. This suggestion finds further confirmation when looking at the oxygen NEXAFS spectra of $La_{1-x}Sr_xFe_{0.75}Ni_{0.25}O_{3-\delta}$ for x=0 and x=0.5, as shown in Figure 3. $La_{1-x}Sr_xFeO_{3-\delta}$ has an $e_g$ spin up orbital symmetry state from doped holes, and a $t_{2g}$ and $e_g$ spin down doublet from hybridized O(2p)-Fe(3d) states as the O 1s NEXAFS pre-edge structure. The intensity of the $e_g$ states in $La_{1-x}Sr_xFeO_{3-\delta}$ scale roughly with the Sr content [13]. Since an exponential dependence between the relative spectral intensity ratio $e_g\uparrow/(t_{2g}\downarrow+e_g\downarrow)$ and the conductivity was found [14], the assessment of these pre-edge structures is relevant for transport related functionality in devices. $La_{1-x}Sr_xFe_{0.75}Ni_{0.25}O_{3-\delta}$ with x=0 has a small such leading $e_g\uparrow$ peak, supporting the aforementioned suggestion that Ni substitution has the same effect like hole doping. It has recently been shown that increasing Ni concentration in $LaFe_{1-x}Ni_xO_3$ increases the intensity of pre-peak and also shifts it towards $E_F$ [15]. The intensity of the corresponding $e_g$ spin up peak in $La_{1-x}Sr_xFe_{0.75}Ni_{0.25}O_{3-\delta}$ for x=0.50 exceeds by far the intensity of that for x=0, i.e. by a factor of six. In addition we do notice a small shift of the of the $e_g$ spin up peak energy position towards $E_F$ upon hole doping. We have recently shown that, since the Ni 3d states are closer to $E_F$ than the Fe 3d states, the doped holes go predominantly to Ni [9]. The changes in the electronic structure of the $La_{1-x}Sr_xFe_{0.75}Ni_{0.25}O_{3-\delta}$ upon Sr substitution are qualitatively reflected by the electrical conductivity, as shown in ref. [9]. The well known conductivity maximum of $La_{1-x}Sr_xFeO_{3-\delta}$ at x = 0.50 is recovered for $La_{1-x}Sr_xFe_{0.75}Ni_{0.25}O_{3-\delta}$ for the same x; this is also corroborated by the $e_g\uparrow$ peak intensity in the oxygen NEXAFS spectra for x = 0 and x = 0.50, Figure 3.

While the influence of the Ni - when to 25% substituting the Fe in $La_{1-x}Sr_xFeO_{3-\delta}$ - on the conductivity and on the electronic structure appears clear now, the mechanism of how the Ni interacts in relation to the Fe with oxygen is not yet clear. Quite interesting, we notice that the intensity for x=0.75 in the VB PES $e_g$ intensity is a little yet noticeable larger than that for x=0.50. The same holds for the larger extent for the $t_{2g}$ intensity, corroborating how also the electrical conductivity depends systematically on the VB PES intensity.

**CONCLUSIONS**

We have performed resonant VB PES, and NEXAFS at O K edge and Fe $L_{2,3}$ edges on $La_{1-x}Sr_xFe_{0.75}Ni_{0.25}O_{3-\delta}$ (x=0.0, 0.5, and 0.75). The spectral weight of occupied $e_g$ and $t_{2g}$ states of Fe is reduced upon electron hole doping up to 50% and then starts to increase which is parallel to changes in electrical conductivity. The spectral weight of the pre-peak in the oxygen NEXAFS spectra due to the p-type electron holes created on Ni 3d increases with increasing x. With respect to $La_{1-x}Sr_xFeO_3$, it is observed that the spectral weight transfers from below $E_F$ to above it across the gap. However, here in $La_{1-x}Sr_xFe_{0.75}Ni_{0.25}O_{3-\delta}$ it is difficult to conclude this since the Ni 3d closer to $E_F$ than Fe 3d and we do not consider any metal (Fe)-metal (Ni) electron hole transfer.

**Acknowledgment**

Funding by E.U. MIRG No. CT-2006-042095, Swiss NSF No. 200021-116688, Swiss Federal Office of Energy project No.100411. The ALS is supported by the Director, Office of Science/BES, of the U.S. DoE, No. DE-AC02-05CH11231.

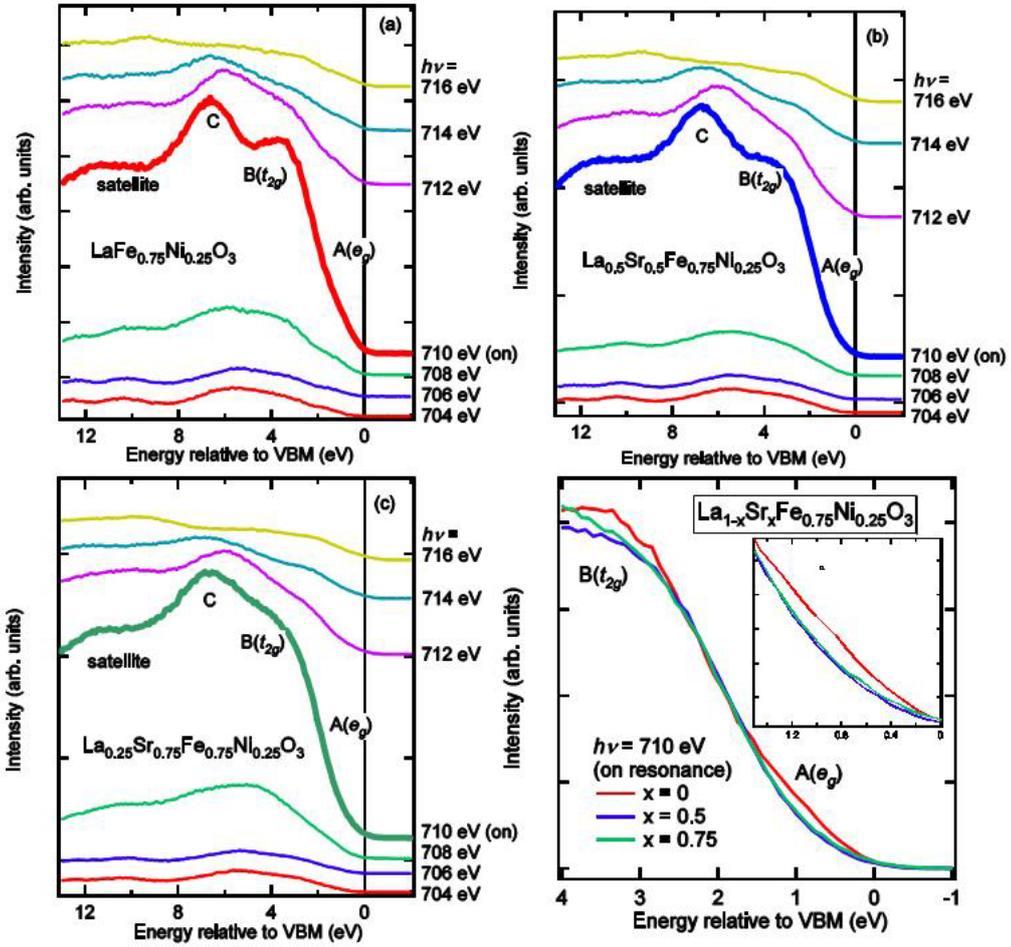

**Figure 1.** Fe 2p–3d resonant valence-band photoemission spectra of $La_{1-x}Sr_xFe_{0.75}Ni_{0.25}O$. There is on-resonance condition at 710 eV. (a) x = 0, (b) x= 0.50, (c) x = 0.75. (d) Expanded valence-band spectra at Fe 2p-3d on-resonance (710 eV), with an inset showing that the intensity for x=0.75 is in between those for x=0 and X=0.5.

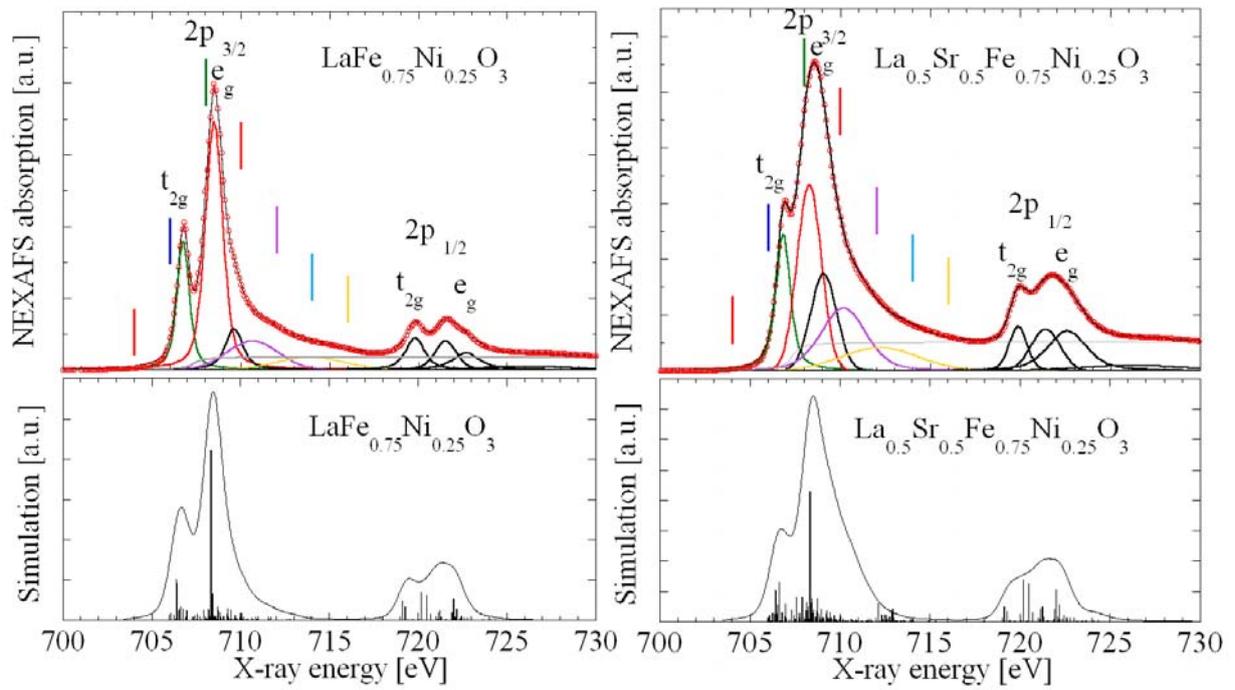

**Figure 2.** Top panels - Normalized Fe $L_{2,3}$ edge x-ray absorption spectra of $La_{1-x}Sr_xFe_{0.75}Ni_{0.25}O$ for x=0, and x=0.50. Vertical lines denote the energies where VB PES spectra were collected. Bottom panels – Atomic multiplet simulation spectra for the same stoichiometries.

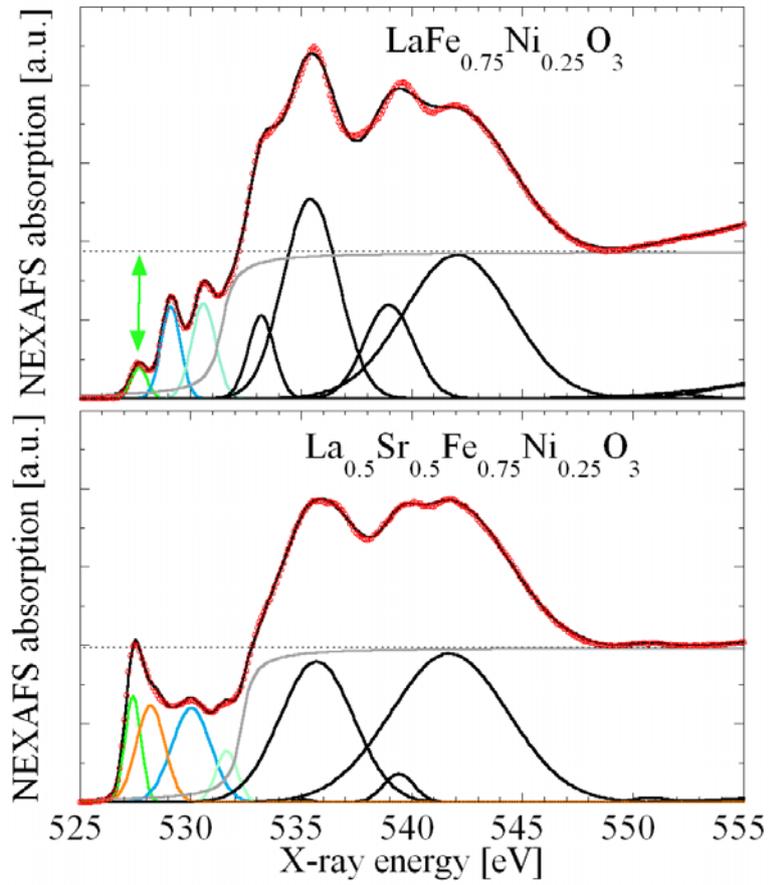

**Figure 3.** Normalized O K edge x-ray absorption spectra of $La_{1-x}Sr_xFe_{0.75}Ni_{0.25}O$ for x=0, and x=0.50 with suggested peak deconvolution into Voigt functions. Arrow in top panel spectrum shows the difference between $e_g\uparrow$ peak intensity and normalization level (dashed horizontal line). Spectrum in bottom panel has additional peak at ~ 528 eV.